# Alpha-particle generation from H-$^{11}$B fusion initiated by laser-accelerated boron ions

**Defeng Kong[1], Shirui Xu[1], Yinren Shou[1], Ying Gao[1], Zhusong Mei[1], Zhuo Pan[1], Zhipeng Liu[1], Zhengxuan Cao[1], Yulan Liang[1], Ziyang Peng[1], Pengjie Wang[1], Di Luo[2,3], Yang Li[2,3], Zhi Li[2,3], Huasheng Xie[2,3], Guoqiang Zhang[4], Wen Luo[5], Jiarui Zhao[1,§], Shiyou Chen[1], Yixing Geng[1], Yanying Zhao[1], Jianming Xue[1], Xueqing Yan[1,6,7], and Wenjun Ma[1,6,7,*]**

[1] State Key Laboratory of Nuclear Physics and Technology, and Key Laboratory of HEDP of the Ministry of Education, CAPT, Peking University, Beijing 100871, China
[2] Hebei Key Laboratory of Compact Fusion, Langfang 065001, China
[3] ENN Science and Technology Development Co., Ltd., Langfang 065001, China
[4] Shanghai Institute of Applied Physics, Chinese Academy of Sciences, Shanghai, 201800, China
[5] School of Nuclear Science and Technology, University of South China, Hengyang 421001, China
[6] Beijing Laser Acceleration Innovation Center, Huairou, Beijing 101400, China
[7] Institute of Guangdong Laser Plasma Technology, Baiyun, Guangzhou 510540, China

§corresponding author: jrzhao@pku.edu.cn
*corresponding author: wenjun.ma@pku.edu.cn

**Abstract**

Here we report the generation of MeV alpha-particles from H-$^{11}$B fusion initiated by laser-accelerated boron ions. Boron ions with maximum energy of 6MeV and fluence of $10^9$/MeV/sr@5MeV were generated from 60-nm-thick self-supporting boron nanofoils irradiated by 1J femtosecond pulses at an intensity of $10^{19}$W/cm$^2$. By bombarding secondary hydrogenous targets with the boron ions, $3\times10^5$/sr alpha-particles from H-$^{11}$B fusion were registered, which is consistent with the theoretical yield calculated from the measured boron energy spectra. Our results demonstrate an alternative way toward ultrashort MeV alpha-particle sources employing compact femtosecond lasers. The ion acceleration and product measurement scheme are referential for the studies on the ion stopping power and cross-section of the H-$^{11}$B reaction in solid or plasma.

## 1. Introduction

Nuclear fusion between proton(H) and boron($^{11}$B), $^{11}_{5}B + p \rightarrow 3\alpha + 8.68 MeV$, is a widely concerned reaction[1-7] due to its appealing potential in fusion energy harness[8-10]. Unlike the D-D reaction and the D-$^6$Li/D-T cycle[11], the H-$^{11}$B reaction releases alpha-particles instead of neutrons, which offers clean energy without neutron radiation hazards. More importantly, $^{11}$B is stable and abundant on earth, which sheds off the fuel problem in D-T fusion. With the rapid development of high-power lasers, laser fusion based on the H-$^{11}$B reaction attracts more and more attention. However, self-sustained H-$^{11}$B fusion under equilibrium conditions is highly challenging due to the insurmountable radiation loss problem at elevated temperatures. Many explorations on the H-$^{11}$B reaction have been ongoing, such as driving the fusion out of thermal equilibrium by using ultrashort lasers[12-14] to reduce the radiation loss or revisiting the fusion reactivity[15-17] in the plasma environment.

In addition to the potential for clean fusion energy, the alpha-particle generation from the H-$^{11}$B reaction could be a valuable source for medical and industrial applications[18-20]. The cross-section for the H-$^{11}$B reaction is very large, e.g., 1.2 barn[21] at 620 keV (center-of-mass energy), and one reaction can release one of 1-MeV and two of 4-MeV alpha-particles [15, 22] in a simplified view. With high-energy reactants, the yield and the kinetic energies of the alpha-particles could be prominent, depending on the reaction channels. The alpha-particle generation from laser-driven H-$^{11}$B reaction was firstly reported in 2005 with a yield of $10^3$/sr/shot[23] using a boron-rich polyethylene target irradiated by a picosecond laser. In subsequent experiments, the yields have been continuously increased to $10^6\alpha$/sr/shot and $10^9\alpha$/sr/shot[12, 24] in the so-called "pitcher-catcher" scheme, where energetic protons are produced from a μm-thick target through target normal sheath acceleration(TNSA) and bombard a secondary boron target.

Besides of the boron-rich polyethylene targets, "sandwich" targets (SiH/B/Si) and thick boron-nitride (BN) targets were irradiated with kilojoule-scale sub-ns lasers, producing $10^9$ and $10^{10}$/sr/shot alpha-particles, respectively[25-27]. It was found that, in spite of the difference in the driving lasers, the observed yield of the alpha-particles had a similar scaling law of about $10^5$-$10^6 \alpha$/sr/J.

Up to now, all the reported alpha-particle generation was driven by low-repetition-rate, high-energy, long-pulse lasers. Operating one-shot typically takes an hour or more, which severely limits potential applications. Routes that employ femtosecond lasers as the drivers are noteworthy to study, which can operate at a much higher repetition rate. Besides the high repetition rate, another advantage of femtosecond lasers is that their intensities are much higher than long-pulse lasers for the given pulse energy. The 100s TW or PW femtosecond lasers can deliver intensities of $10^{18}$-$10^{22}$W/cm$^2$ on the targets. Laser ion acceleration at such high intensity can produce copious MeV ions from non-equilibrium laser-plasma interaction, matching the cross-section's apex nicely.

Moreover, all the reported studies of laser-ion-initiated H-$^{11}$B fusion utilize protons to bombard boron targets[12, 23-28]. If the opposite scheme, i.e., initiating H-$^{11}$B fusion with energetic boron ions, is adopted, the generated alpha-particles would be more directional due to the higher mass of boron atoms[29]. The yield may also be enhanced as studies show that the energy conversion efficiencies from laser energy to heavy ions are higher than that to protons in favorable acceleration regimes[30, 31]. Furthermore, this scheme can be employed to investigate the stopping power of boron ions inside solid or plasma targets, which is very important for future H-$^{11}$B nuclear reactors [29, 32, 33]. However, the alpha-particle generation by bombarding hydrogenous solid or plasma targets with laser-accelerated boron ions has not been realized yet. The main reason is the shortage of energetic laser-accelerated boron ions. In the widely adopted TNSA regime, the targets are $\mu$m-thick solid foils[34]. Ions with the highest charge to mass ratio, i.e., protons, favorably gain energy from the sheath field. The acceleration of heavy ions is drastically suppressed. With the development of laser and target-fabrication technology, ultrathin targets with nm-scale thickness were allowed to be used in the experiments, indicating the prominent efficiency for heavy-ion acceleration. The variation of the laser and target parameters leads to different regimes such as radiation pressure acceleration(RPA)[35, 36], relativistic induced transparency(RIT)[37-39], breakout afterburner acceleration(BOA)[40, 41], or hybrid acceleration[31, 42]. So far, energetic heavy ions such as C$^{6+}$, Al$^{13+}$, and Au$^{51+}$ have been produced with maximum energy up to 1.2 GeV[31, 39, 43].

In this work, we report the first H-$^{11}$B fusion and alpha-particle generation results by bombarding hydrogenous targets with laser-accelerated boron ions. MeV-level boron ions were accelerated from 60-nm-thick boron targets under the irradiation of high-contrast femtosecond laser pulses. The alpha-particles from H-$^{11}$B fusion were measured by CR39 ion track detectors. The fusion reactions happening inside the hydrogenous targets are discussed considering the ion-nuclear collision, and the theoretical yield is calculated based on the measured $^{11}$B spectra, which is consistent with our experimental results.

## 2. Experimental setup

### 2.1 Laser parameters

The experiment was performed on the 200TW CLAPA Ti:sapphire laser system at Peking University[44]. The experimental layout is shown in fig.1(a). An s-polarized laser pulse was normally focused on the 60-nm-thick boron nanofoils with the spot size of 8.4×9.2$\mu$m (full width at half maximum) by an f/3 off-axis-parabolic mirror. The central wavelength and duration of the laser pulse were 800nm and 30fs, respectively. A cross-polarized wave system and a single plasma mirror system were employed to improve the laser contrast ratio up to $10^9$@40ps and prevent the damage of targets from pre-pulses. The on-target laser energy was 1J, corresponding to a peak intensity of $1\times10^{19}$ W/cm$^2$. A 5-$\mu$m-thick plastic (C$_{10}$H$_8$O$_4$) foil with the proton density of $4\times10^{22}$/cm$^3$ was located 0.5 mm behind the targets at the laser axis as the "catcher" for H-$^{11}$B reactions.

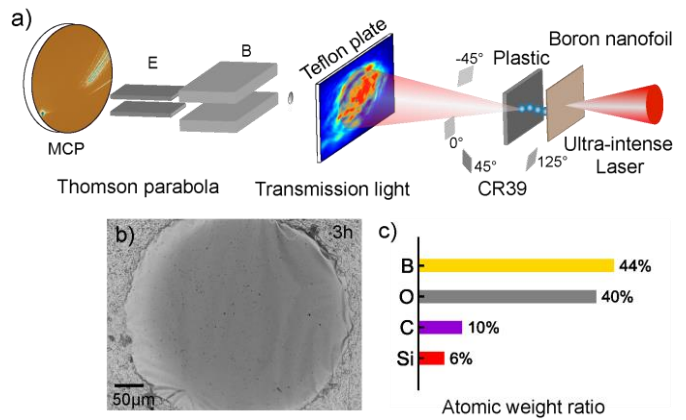

**Figure 1.** a) Experimental setup. The laser pulses irradiate a boron nanofoil with normal incidence. A 5-$\mu$m-thick plastic foil was located 0.5 mm behind the boron nanofoil to initiate the H-$^{11}$B fusion. The CR39, TPS, and Teflon plate were placed around the target to measure the alpha-particles, boron ions and collect the transmitted light, respectively. b) Top-view morphology of self-supporting nanofoils three hours after preparation c) Atomic weight ratio of the boron nanofoil.

### 2.2 Boron nanofoil target

The boron nanofoils were prepared by the RF-magnetron sputtering deposition using the natural boron. The atomic ratio of $^{10}$B and $^{11}$B is 1:4. The details of the target fabrication method will be reported elsewhere. To optimize the ion acceleration, we used 60-nm-thick self-supporting B foils, the thinnest that could be fabricated at that time, as the targets in the experimental campaign. Fig.1(b) depicts the top-view morphology of a 60-nm-thick self-supporting boron nanofoils on a target hole with a diameter of 0.5mm. The chemical composition of the targets is characterized by an energy dispersive spectrometer in a scanning electron microscope (fig.1(c)). Due to the oxidation of the targets in the air, the atomic ratio of B:O is 1.1:1. Besides, The Si atoms are from the silicon wafer as a target substrate during the fabrication. The C atoms come from the contaminated layer of nanofoils. The density of the foils, measured by the weighting method, is about 0.95g/cm$^3$. If the target is fully ionized, the electron density would be $n = 160n_c$, here the critical density $n_c = m_e\omega^2/(4\pi e^2) = 1.7 \times 10^{21} cm^{-3}$.

## 2.3 Diagnostics

The energy spectra of the ions were measured by a Thomson Parabola Spectrometer(TPS) with a microchannel plate(MCP) equipped with a phosphor screen positioned 0.78 m away from the targets in the normal direction of the targets. The collimated ions with different energy and charge-to-mass ratio(CMR) were deflected by the electromagnetic fields and hit on the MCP with parabola traces. Ion signal multiplied by the MCP was converted to optical signals captured by a 16-bit EMCCD camera. For a good resolution of the traces, a tiny collimating aperture was employed. The corresponding acceptance angle is only $4.2\times10^{-8}$sr, which allows the recognition of single-ion events on the MCP[43, 45]. A Teflon plate with a through-hole was placed behind the target to collect the transmitted light, which can be used as a diagnostic for the laser-plasma interaction.

The alpha-particles generated from H-$^{11}$B fusion were detected by CR39 ion track detectors at angles of -45°, 0°, 45°, and 125°. Here 0° is the laser axis direction. The distance between CR39 and the targets was 130mm. The CR39 sheets were wrapped in 10$\mu$m-thick aluminum foils to block low-energy ions. According to the Monte Carlo simulation results from SRIM[46], the minimum energy required to penetrate 10$\mu$m aluminium for proton, alpha-particle, boron, carbon, and oxygen ions is 0.8 MeV, 2.9 MeV, 9.5 MeV, 12 MeV and 16.5 MeV, respectively. In our experiments, all the carbon and oxygen ions were blocked by the Al foils (see below), and only protons and alpha-particles with energy above 0.8 MeV and 2.9 MeV can go through and result in visible traces in CR39 after etching.

## 3. Result

### 3.1 Energy spectra of borons and other ions

The absolute energy spectra of boron ions can be obtained from our TPS. Fig.2(a) shows a raw image recorded by TPS after the shooting (without the secondary plastic foil). More than ten spectral lines from boron, carbon, oxygen ions and protons can be identified. The parabolic traces of $^{11}$B ions are marked with different lines. Boron ions with high charge states ($^{11}$B$^{3+}$, $^{11}$B$^{4+}$, $^{11}$B$^{5+}$) can be clearly identified. Different from protons and carbon ions, the traces of boron ions are composed of cluster signals with similar shapes and clear boundaries. Due to the small acceptance angle of the TPS, the boron ions are sparsely distributed on the parabolic traces, and a distinct cluster signal is the response of a single boron ion hitting in MCP, indicating a 'single-ion' event. By summing the counts for distinct clusters as the function of ion energy, we can obtain the response of a single boron ion[43]. Based on the single-ion response data, the absolute energy spectra of $^{11}$B$^{5+}$, $^{11}$B$^{4+}$, and $^{11}$B$^{3+}$ ions have been derived in fig.2(b). The vertical error bars come from deviations of the single-ion response, and horizontal error bars reflect the width of the energy bins, which was adopted to 0.2 MeV to obtain smooth spectra curves. We can find that the maximum energy of $^{11}$B$^{3+}$, $^{11}$B$^{4+}$, and $^{11}$B$^{5+}$ is 2.7 MeV, 4.2 MeV, and 5.8 MeV, respectively. The corresponding ion temperature is 0.25 MeV, 0.25 MeV, and 0.47 MeV, respectively. The typical fluence is 10$^8$-10$^{10}$/MeV/sr, depending on the energy. For instance, the fluence of $^{11}$B$^{5+}$ is 10$^9$/MeV/sr at 5MeV. The spectra of proton, carbon, and oxygen ions from this shot are given in fig.2(c).

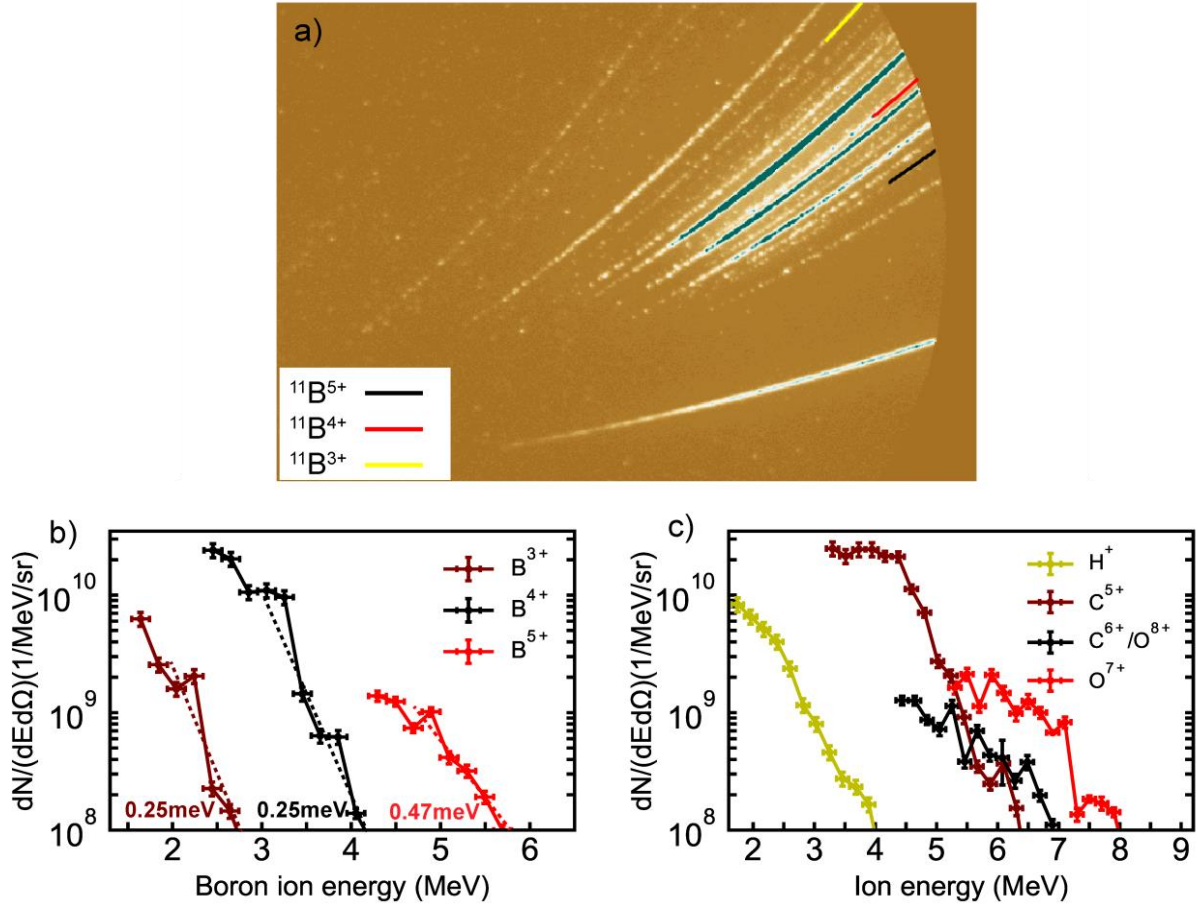

**Figure 2.** a) Raw TPS data from a 60-nm-thick boron nanofoil. The parabolic traces of $^{11}B^{3+}$, $^{11}B^{4+}$ and $^{11}B^{5+}$ ions have been marked with different lines. b) Ion spectra of $^{11}B^{3+}$, $^{11}B^{4+}$ and $^{11}B^{5+}$. c) Ion spectra of proton, carbon, and oxygen ions.

*3.2 Alpha-particle measurement*

The CR39 sheets used for alpha-particle measurement were etched in 6 mol/L NaOH solution at 98°C for 2 hours to reveal the ion tracks. Fig.3(a) displays the CR39 images with a solid angle of $3.3\times10^{-5}$sr at the angle of 0° and 125° after three shots in a row. A control CR39 sheet that was not put inside the chamber was also etched with the same procedure, whose surface morphology was shown in fig.3(a) as well.

According to fig.2(b-c), the maximum energy of laser-accelerated boron, carbon, and oxygen ions is 6MeV, 7MeV, and 8MeV, respectively. Therefore, those ions were completely blocked by the Al foils. The tracks of protons and alpha-particles can be easily distinguished from each other based on their sizes. We referred to the calibration of protons and alpha-particles from Y. Zhang et al.'s work under the same etch condition[47], shown as the lines in fig.3(b). Therefore, the dense grey dots with diameters of 4-6 $\mu$m represent the protons, while the alpha-particles are larger black pits with diameters of 20-30$\mu$m in fig.3(a). According to the proton's spectrum from TPS and considering their energy loss in the plastic and Al foil, the proton tracks in the CR39s can be estimated as $9\times10^9$/sr at the 0°. So about $10^5$ protons should be observed in the CR39 image within a solid angle of $3.3\times10^{-5}$sr, consistent with the high number density of grey dots. We can find about 31 alpha-particle tracks at the 0° direction, and only 7 at 125°. The energy of the alpha-particles can be roughly estimated from the size of the tracks. The brown circles in fig.3(b) show some representative alpha-particles from 0° direction. The energy range of alpha-particles is 3-5MeV, which is consistent with the kinetic energy obtained from the fusions reaction. By counting the number of the alpha-particles, we can get the averaged angular distribution of alpha-particle flux per shot in fig.3(c). Due to the off-line measurement of CR39 and the limited beamtime, we did not perform more shots and, unfortunately, can not give the shot-to-shot fluctuations. Generally speaking, the angular distribution shows a directional feature in the forward direction due to the momentum of the boron ions. The peak yield is $3\pm0.2\times10^5$/sr/J, and the experimental uncertainty comes from the statistical error of tracks on CR39. It should be noted that the given values in fig.3(c) are conservative as only alpha-particles with energies above 2.9MeV can be detected after the shielding of the Al foils.

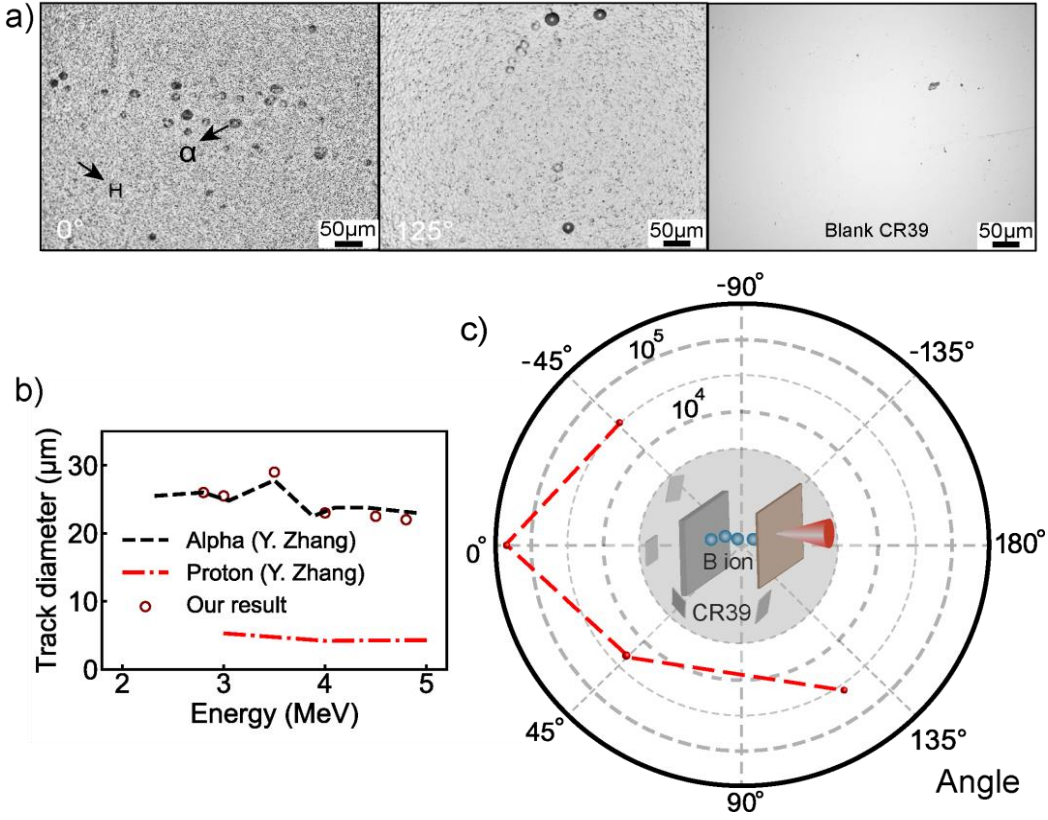

**Figure 3.** Alpha-particle generation from H-$^{11}$B fusion measured by CR39. a) Raw images of CR39 sheets. b) Calibrated track diameters versus the energy of protons and alpha-particles[47] and representative alpha-particles of our result. c) Angular dependence of alpha-particle flux. The inset shows the experimental layout of the pitcher-catcher scheme.

## 4. Discussion

We can theoretically calculate the yield of the alpha-particles from the measured boron spectra and compare it with that from the CR39 measurement. The number of boron-induced fusion reactions $N_f$ can be estimated using the differential equation describing the ion-nuclear collisional process in the target nucleus[48] with a thickness of $D$ as

$$dN_f = \int n\sigma(E)v_b(x)dt dN_b = n\int_0^D \sigma(E)dx \frac{dN_b}{dE^*}dE^*. \quad (1)$$

where $n = 4\times 10^{22}$/cm$^3$ is the proton density of the target nucleus, $\sigma(E)$ is the nuclear cross-section and $v_b, N_b$ are the velocity and number of incident ions, respectively. $dN_f$ is the number of reactions driven by the boron ions impinging on the target with kinetic energy between $E^*$ and $E^* + dE^*$. $\frac{dN_b}{dE^*}$ represents the energy spectrum of the incident boron ions, reported in fig. 2(b). The thickness $D$ of 5μm is close to the projected range $R_{E=5.8\text{MeV}} = 7.8\mu m$ for boron ions with the maximum energy of 5.8MeV. Although some high-energy boron ions can pass through the second target, the number is small, an order of magnitude lower than that of the 4MeV-boron ions shown in fig.2(b). Moreover, their kinetic energy has degraded to below 2MeV, corresponding to a pretty low fusion reactivity. Therefore, we believe that most boron ions are exhausted and stopped in the target nucleus for simplicity. The Eq. (1) can be further expressed in terms of the energy $E$ of boron ions,

$$dN_f = n\int_{E^*}^0 \frac{\sigma(E)}{dE/dx}dE \frac{dN_b}{dE^*}dE^* = n\int_0^{E^*} \frac{\sigma(E)}{S(E)}dE \frac{dN_b}{dE^*}dE^*. \quad (2)$$

where $S(E) = -dE/dx$ represents the stopping power of the target nucleus against incident boron ions. By integrating the energy $E$, the number of all alpha-particles generated from H-$^{11}$B fusion can be expressed as

$$N_\alpha = 3N_f = 3n\int_0^{E_0}\left(\int_0^{E^*}\frac{\sigma(E)}{S(E)}dE\right)\frac{dN_b}{dE^*}dE^*. \quad (3)$$

Fig.4 depicts the curves of $\sigma(E)$, $S(E)$, $\frac{dN_b}{dE^*}$ as the function of boron-ion energy. The $S(E)$ in plastic (C$_{10}$H$_8$O$_4$) is simulated with SRIM[46], including the electronic and nuclear energy loss based on the cold target. The $\sigma(E)$ of H-$^{11}$B fusion is expressed according to Nevins and Swain's results[21, 49] and polynomially fitted as given in table 1. The $\frac{dN_b}{dE^*}$ of $^{11}$B$^{5+}$, $^{11}$B$^{4+}$ and $^{11}$B$^{3+}$ are also exponentially fitted in table 1 according to fig.2(b). The low-energy boron ions that were not measured by the TPS are also included by extrapolation down to 1MeV. Table 1 gives the theoretical yield of alpha-particle from $^{11}$B$^{5+}$, $^{11}$B$^{4+}$ and $^{11}$B$^{3+}$ ions. One can find that the contribution from $^{11}B^{5+}$ and $^{11}B^{4+}$ is 64.9% and 34.8%, respectively. The energy of contribution from $^{11}B^{4+}$ can not be ignored even though their energy is lower than that of $^{11}B^{5+}$. The total yield of alpha-particles is $N_\alpha \approx 1.6 \times 10^5$/sr, which matches the experimental measurement from CR-39 very well. Besides the $^{11}B(p,\alpha)2\alpha$, other channels such as $^{12}C(p,\alpha)$ and $^{16}O(p,\alpha)$ can also contribute to the alpha-particles generation. However, the cross-section of these reactions is two to three orders of magnitude lower at the relevant energy[50]. Based on the measured energy spectra, the estimated total alpha-particles yield from the accelerated C, O, and H is about $10^3$/sr, two orders of magnitude lower than the observation.

The theoretical and measured alpha-particle yield of $10^5 \alpha$/sr/shot with 1J femtosecond laser pulses reaches a similar level to the case of proton-induced laser-driven fusion[12, 24, 25]. It should be noted that the reaction condition is still far from the apex of the cross-section at 7.4 MeV($\sigma = 1.2$barn) (see fig.4). Futher enhancement of the energy of B ions would lead to a higher yield and better collimation of the alpha-particles. Our simulation shows that the yield can be increased from $1.04 \times 10^5$ to $1.89 \times 10^6 \alpha$/sr/shot if the maximum $^{11}B^{5+}$ energy and the temperature can be enhanced to 13MeV and 1.3MeV, respectively (total $^{11}B^{5+}$ ion number keeps the same), which is very promising at higher laser intensities. Alternatively, further optimizing the thickness of the targets would also lead to higher ion energy and more H-$^{11}$B reactions. Our 60-nm-thick B targets are slightly thicker than the optimum thickness of 10nm in the RPA regime, according to $l \sim \lambda a_0 n_c/n_e$. Here $a_0 = 2.2$ is the normalized laser field. In these targets, TNSA probably is the primary acceleration mechanism. Further reducing the thicknesses of the targets would enable us to utilize more favorable acceleration regimes such as RPA or RIT.

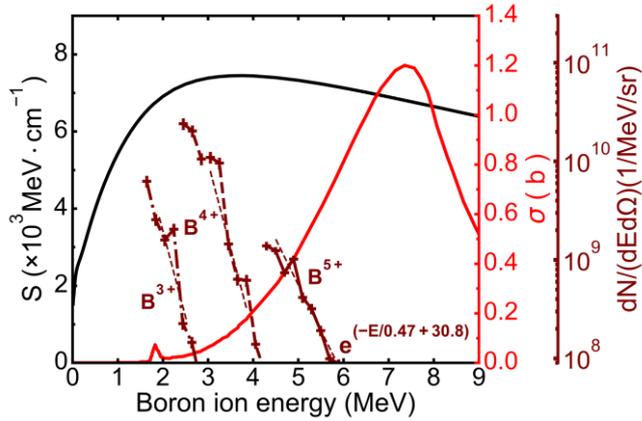

**Figure 4.** Blackline: the stopping power $S(E)$ of boron ions. Redline: the cross-section $\sigma(E)$ of H-$^{11}$B fusion as a function of boron-ion energy in the lab. Brown lines: the energy spectra $\frac{dN_b}{dE^*}$ of $^{11}B^{5+}$, $^{11}B^{4+}$ and $^{11}B^{3+}$. The dashed lines are the exponential fitting of energy spectra.

**Table 1.** Curve fitting functions of $\sigma(E)$, $S(E)$ and $\frac{dN_b}{dE^*}$ and the number of alpha-particles $N_\alpha$.

| Ions | $^{11}B^{5+}$ | $^{11}B^{4+}$ | $^{11}B^{3+}$ | Total |
|---|---|---|---|---|
| $\frac{dN_b}{dE^*}$ | $e^{(-E/0.47+30.8)}$ | $e^{(-E/0.25+34.3)}$ | $e^{(-E/0.25+29.2)}$ | -- |
| $\sigma(E)$ | $0.0004934E^5 - 0.0132E^4 + 0.1155E^3 - 0.3694E^2 + 0.4366E - 0.115$ | | | -- |
| $S(E)$ | $0.0006076E^5 - 0.02117E^4 + 0.2801E^3 - 1.767E^2 + 5.194E + 1.748$ | | | -- |
| $N_\alpha$ | $1.04 \times 10^5$ | $5.58 \times 10^4$ | $3.78 \times 10^2$ | $1.6 \times 10^5$ |
| ratio | 64.9% | 34.8% | 0.3% | 100% |

† The units of the parameters match the axes in fig.4.

## 5. Conclusion

In summary, we report the generation of $3 \pm 0.2 \times 10^5$/sr/J alpha-particles initiated by boron ions driven by a compact femtosecond laser for the first time. The yield is in good agreement with the theoretical calculation based on the measured $^{11}$B spectra, the stopping power of the boron ions in solid targets, and the reported cross-section of H-$^{11}$B fusion. Our results demonstrate an alternative way toward ultrashort MeV alpha-particle sources with compact femtosecond lasers. The ion acceleration and product measurement scheme can provide a referential method for future studies on the stopping power of boron ions and the corresponding nuclear cross-section of H-$^{11}$B fusion in plasma by heating the "catcher" target into plasma. With higher laser intensities or thinner nanofoils in the future, the energies and number of boron ions would further increase. The resulting higher-yield and directional alpha-particles at high-repetition-rate could be promising for medical studies and industrial applications.

### Acknowledgments

This work was supported by the following projects: NSFC innovation group project [grant number 11921006] and National Grand Instrument Project [grant number 2019YFF01014402].